# Sign-Reversal Coupling in Coupled-Resonator Optical Waveguide

Zhen Gao[†1], Fei Gao[†1], Youming Zhang[1], and Baile Zhang*[1,2]

[1]*Division of Physics and Applied Physics, School of Physical and Mathematical Sciences, Nanyang Technological University, Singapore 637371, Singapore.*

[2]*Centre for Disruptive Photonic Technologies, Nanyang Technological University, Singapore 637371, Singapore.*

[†]*These two authors contributed equally to the work.*

[*]*Authors to whom correspondence should be addressed; E-mail: blzhang@ntu.edu.sg (B. Zhang)*

# Abstract

**Coupled-resonator optical waveguides (CROWs), which play a significant role in modern photonics, achieve waveguiding through near-field coupling between tightly localized resonators. The coupling factor, a critical parameter in CROW theory, determines the coupling strength between two resonators and the waveguiding dispersion of a CROW. However, the original CROW theory proposed by Yariv *et al*. only demonstrated one value of coupling factor for a multipole resonance mode. Here, by imaging the tight-binding Bloch waves on a CROW consisting of designer-surface-plasmon resonators in the microwave regime, we demonstrate that the coupling factor in the CROW theory can reverse its sign for a multipole resonance mode. This determines two different waveguiding dispersion curves in the same frequency range, experimentally confirmed by matching Bloch wavevectors and frequencies in the CROW. Our study supplements and extends the original CROW theory, and may find novel use in functional photonic systems.**

# Significance Statement

**Coupled-resonator optical waveguides (CROWs), as originally proposed by Yariv et al., play a significant role in modern photonics. However, the original CROW theory demonstrated only one value of coupling factor for a multipole resonance mode. Here, we theoretically and experimentally demonstrate that the coupling factor in the CROW theory for a multipole resonance mode can reverse its sign. This sign-reversal coupling determines two types of waveguiding mechanisms on a CROW in the same frequency range, experimentally confirmed by matching Bloch wavevectors and frequencies in the CROW. As a fundamental supplement and extension of the original CROW theory, our study may find novel use in functional photonic systems.**

Coupled-resonator optical waveguides (CROWs) (1,2,3), as originally proposed by Yariv *et al.* (1), are a classic textbook model in photonics (4,5) with various applications, ranging from slow-light engineering (6-12), quantum simulation (13,14), to most recently topological photonics (15-18) and $\mathcal{PT}$-symmetric photonics (19,20,21). Underlying these wide physics and engineering applications is the near-field coupling between adjacent tightly localized high-$Q$ resonators, whose phenomenon is analogous to the tight-binding model of atomic wave functions in a solid-state lattice (22). A critical parameter in CROW theory, the coupling factor (1), arises from the overlap between resonance wave fields of two coupled resonators. This factor not only measures the coupling strength between the two coupled resonators, but also determines the waveguiding dispersion of CROW. However, in Yariv's original CROW theory, only one value of coupling factor has been assumed for a multipole resonance mode (1).

Here, on the platform of two-dimensional (2D) designer-surface-plasmon (DSP) metamaterials (23,24,25), we directly image the Bloch waves on a one-dimensional (1D) CROW that consists of a sequence of DSP resonators (termed as "meta-atoms" or "metamaterial particles" in the field of metamaterials) (24,25,26). By comparing the measured phase relation between two coupled DSP resonators, we demonstrate that the coupling factor can reverse its sign for a single multipole resonance mode. We further measure the Bloch waves on the CROW and discern two different dispersion curves in the same frequency range, corresponding to the two opposite coupling factors. Finally, we construct a right-angle sharp bend of CROW in which the sign reversal of coupling factor occurs spatially at the bending corner.

In Yariv's original CROW theory (1), the coupling between two identical resonators, as shown in Fig. 1*A*, is through overlapping two "poles," or intensity maxima, of resonance wave patterns that are tightly confined around the two resonators. Following Yariv's analysis (1), we assume, for simplicity, that the 2D resonators are made of homogeneous and isotropic material with constant permittivity. Defining the *x*-axis as in Fig. 1A and setting the radius of two identical resonators as $R$, the multipole resonance mode profile of each resonator can be expressed as $E = B_m(kr)\cos(m\varphi)$, where $m$ is the order of multipole resonance mode, and $(r, \varphi)$ are the polar coordinates with the center of each resonator as the origin. $B_m(kr) = \begin{cases} J_m(k_d r), & r < R \\ \frac{J_m(k_d R)}{H_m^1(ik_0 R)} H_m^1(ik_0 r), & r > R \end{cases}$, where $J_m(.)$ is Bessel function, $H_m^1(.)$ is first-kind Hankel function, and $k_d$ and $k_0$ are wavevectors in resonators and the environment, respectively. The coupling factor $\kappa$ is a mathematical expression that depends on the integration of overlapped resonance wave fields of the two resonators (1). (See Supplementary Information for the mathematical expression.) This kind of "pole-pole" coupling is understandable, because the two intensity maxima facing each other must induce significant coupling between the two resonators.

Interestingly, there is another possibility to couple these two resonators through overlapping two "nodes," or intensity minima, of resonance wave patterns between the two resonators, as shown in Fig. 1B. This "node-node" coupling configuration remains experimentally unobserved in reality. In this case, the multipole resonance mode profile of each resonator can be expressed as $E = B_m(kr)\sin(m\varphi)$. Intuitively, the strength of coupling, or the magnitude of coupling factor $\kappa$, in this "node-node" coupling configuration should be significantly smaller than that in the "pole-pole" coupling configuration. However, a strict derivation shows that the coupling factor $\kappa$ still maintains its magnitude, but simply reverses its sign. (See Supplementary Information for detailed derivation.) This sign-reversal coupling will reverse the phase relation between the two coupled resonators, as we will demonstrate later.

To facilitate the experimental retrieval of coupling factor $\kappa$, we first describe the two coupled resonators with the coupled mode theory (27) as follows:

$$\begin{cases} -\frac{da_1}{dt} = i\omega_0 a_1 + i\kappa\omega_0 a_2 \\ -\frac{da_2}{dt} = i\omega_0 a_2 + i\kappa\omega_0 a_1 \end{cases} \quad (1)$$

where $a_1$ and $a_2$ represent resonance wave fields in the two resonators, and $\omega_0$ denotes the intrinsic resonance frequency of a multipole resonance mode in a single resonator. By solving this eigenvalue problem in Eq. (1), we obtain two orthogonal eigen solutions of $[a_1 \ a_2]^T$ as $[1 \ -1]^T$ and $[1 \ 1]^T$, in which the former corresponds to the out-of-phase ($a_1$ and $a_2$ differ by a phase of $\pi$) coupled mode with eigen frequency $\omega_{1,-1} = \omega_0 - \kappa\omega_0$, and the latter to the in-phase ($a_1$ and $a_2$ have the same phase) coupled mode with eigen frequency $\omega_{1,1} = \omega_0 + \kappa\omega_0$. It can be clearly seen that the magnitude of coupling factor $\kappa$ determines the frequency difference in mode splitting, and the sign of $\kappa$ corresponds to the phase relation of two resonators, i.e. whether they are out-of-phase or in-phase.

We then proceed to experimental demonstration. We adopt recently proposed DSP resonators (25,26) as shown in Fig. 2*A*, which are constructed by decorating periodic subwavelength grooves on circular metallic disks. DSPs are tightly confined electromagnetic modes on subwavelength patterned metallic structures (23), whose dispersion properties and spatial confinements are akin to those of natural surface plasmons at a metal/dielectric interface at optical frequencies. Recently, it has been found that DSP resonators are capable of mimicking localized surface plasmons (24,25,26), opening an opportunity to image directly the near-field coupling mechanism. Because of the analogy between localized electromagnetic resonance modes and atomic wave functions, these DSP resonators are termed as a type of "meta-atoms" (24,25,26) in the field of metamaterials. Another benefit of this DSP-resonator metamaterial platform is that it allows selective excitation of multipole resonance modes, meaning that the analogous "atomic wave functions" can be selectively controlled. We construct a CROW that consists of a sequence of DSP resonators, as shown in Fig. 2*B*. A monopole source is employed at one side of the CROW to selectively excite a multipole resonance mode. The Bloch waves on the CROW is recorded by a near-field probe scanning above the CROW, connected to a microwave network analyzer.

We first consider coupling between two DSP resonators. Figure 3*A* shows the measured near-field transmission spectra through this coupled-resonator dimer. We adopt two excitation configurations. Firstly, we locate the source and the probe in symmetric positions at two ends of the dimer, as indicated by a pair of red dots in the inset of Fig. 3*A*. Secondly, the source and the probe are both placed by one side of two resonators, as indicated by a pair of blue dots in the inset of Fig. 3*A*. We call the first configuration as "end excitation" and the second as "side excitation." Each of these excitation configurations will selectively excite one of the multiple resonance modes. The transmission spectrum for a single resonator is also measured for comparison by placing the source and probe at opposite sides of the resonator. Three resonance modes in a single resonator at 5.34 GHz, 6.07 GHz, and 6.44 GHz can be seen in Fig. 3*A*. They correspond to the quadrupole (labeled as "Q"), hexapole (labeled as "H"), and octopole (labeled as "O") modes, as will be demonstrated later.

Let us consider the end excitation. When the coupled-resonator dimer is excited with end excitation, each of the three resonance modes in a single resonator splits into two supermodes: from one mode at Q=5.34 GHz to two modes at $Q_\pi$=5.19 GHz and $Q_0$=5.49 GHz, from one at H=6.07 GHz to two at $H_0$=6.03 GHz and $H_\pi$=6.11 GHz, and from one at O=6.44 GHz to two at $O_\pi$=6.42 GHz and $O_0$=6.47 GHz. Here the subscript

"π" or "0" denotes out-of-phase or in-phase phase relation for the two resonators after mode splitting. To explicitly show the phase relation, we measure the resonance wave pattern for each resonance mode by scanning the probe in a transverse plane 1 mm above the coupled-resonator dimer, as shown in Fig. 3*B*. The source position is indicated by a red arrow. We observe that for the coupled quadrupole and octopole resonance modes, the two resonators are out-of-phase at lower resonance frequencies ($Q_\pi$ and $O_\pi$), and in phase at higher resonance frequencies ($Q_0$ and $O_0$). However, the situation is reversed for the hexapole mode: the two resonators are in-phase at the lower resonance frequency ($H_0$), and out-of-phase at the higher resonance frequency ($H_\pi$). The different phase relation of two coupled resonators for the coupled hexapole resonance mode is simply because the hexapole mode is an odd mode (mode order *m* =3) while both the quadrupole mode (mode order *m*=2) and the octopole mode (mode order *m*=4) are even modes.

Using the eigen-frequency analysis from Eq. (1), we can experimentally retrieve the coupling factor $\kappa$ from the measured split resonance frequencies and the phase relation between the coupled two resonators. We obtain that with end excitation, $\kappa_Q = 0.0281$, $\kappa_H = -0.0066$, and $\kappa_O = 0.0039$ for the coupled quadrupole, hexapole, and octopole resonance modes, respectively. It is apparent that a higher-order multipole resonance mode has a smaller magnitude of coupling factor due to its tighter field confinement. The negative sign of $\kappa_H$ for the hexapole resonance mode arises from its phase relation of two coupled resonators, which is opposite to that in the quadrupole and octopole resonance modes.

Now we consider the side excitation. It can be seen in Fig. 3*A* that the split resonance frequencies match well with those in the end excitation. For the quadrupole and octopole modes, because of their four-fold rotational symmetry, the resonance wave patterns for their split modes are almost the same with those in the end excitation. We thus do not discuss them here. However, it is not the same situation when it comes to the hexapole mode. As shown in Fig. 3B, we measure the resonance wave patterns for the two split hexapole modes ($H_\pi^*$=6.03 GHz and $H_0^*$=6.11 GHz) in the coupled-resonator dimer, where the source position is indicated by a red arrow. While all previous resonance wave patterns correspond to the "pole-pole" coupling configuration as in Fig. 1A, here these two resonance wave patterns correspond to the "node-node" coupling configuration as in Fig. 1B. Although frequencies of the two split hexapole modes are consistent with those in the "pole-pole" coupling configuration, their phase relation has been reversed: the in-phase mode ($H_0^*$) now has a higher resonant frequency than the out-of-phase mode ($H_\pi^*$). Following the process above to retrieve the coupling factor, we can get $\kappa_H^* = 0.0066$ for the hexapole mode in the "node-node" coupling configuration. This confirms the sign reversal of coupling factor compared to the previously retrieved $\kappa_H = -0.0066$ for the hexapole mode in the "pole-pole" coupling configuration.

We can proceed to construct a CROW with an array of DSP resonators, as shown in Fig. 4A. Each multipole resonance mode in a single resonator now spreads continuously to form a transmission band on the CROW. We describe the CROW that consists of an infinite chain of DSP resonators with coupled mode theory (27) as follows:

$$-\frac{da_n}{dt} = i\omega_0 a_n + i\kappa\omega_0 a_{n-1} + i\kappa\omega_0 a_{n+1} \qquad (2)$$

where $a_n$, $a_{n-1}$ and $a_{n+1}$ denote resonance fields in the *n*-th, (*n*-1)-th and (*n*+1)-th resonators, respectively. The periodicity of this structure allows the application of Bloch theorem which gives $a_{n+1} = a_n\, e^{iK\Lambda}$, where *K* represents the wavevector of Bloch waves on the CROW, and *Λ* is the periodicity of DSP resonators. Then the intrinsic dispersion relation of the Bloch waves on the CROW can be obtained as $\omega = \omega_0\{1+2\kappa\cos(K\Lambda)\}$, being consistent with Yariv's original derivation (1,9). Apparently, the sign of coupling factor $\kappa$ affects the relation between Bloch wavevectors and frequencies.

Given the experimentally retrieved coupling factors, we plot with solid lines in Fig. 4*B* the dispersion curves for the quadrupole, hexapole, and octopole resonance modes along the CROW in the "pole-pole" coupling configuration. This coupling configuration can be achieved with the end excitation, by locating the source and probe at positions as indicated by a pair of red dots in Fig. 4*A*. Working bandwidth gets narrower as the mode order gets higher, because the tighter field confinement in a higher-order mode leads to a smaller magnitude of $\kappa$. Because of the positive sign of $\kappa_Q$ and $\kappa_O$, the dispersion curves of Bloch waves for the quadrupole and octopole resonance modes exhibit properties of "backward waves," in which the phase velocity and the group velocity are in opposite directions. Regarding the hexapole resonance mode, because of its negative coupling factor $\kappa_H$, its Bloch waves are "forward waves," whose phase velocity and group velocity are in the same direction.

Similar to previous discussions on the coupled-resonator dimer, the side excitation in the CROW will turn on the "node-node" coupling configuration for the hexapole resonance mode with a sign-reversal coupling factor, i.e. $\kappa_H^* = -\kappa_H$. The setup of end excitation is indicated by a pair of blue dots in Fig. 4*A*. We plot the dispersion curve for the hexapole resonance mode in the "node-node" coupling configuration in Fig. 4*B* with a dashed red line. This dispersion curve shares the same bandwidth with that of the hexapole resonance mode in the "pole-pole" coupling configuration. Because of the sign reversal of coupling factor, Bloch waves of this dispersion curve in the "node-node" coupling configuration exhibit properties of "backward waves," where the phase velocity and group velocity are in opposite directions.

We adopt two steps in experiment to demonstrate the above predictions. In the first step, we measure transmission spectra to verify transmission bandwidth. The results are shown in Fig. 4*C*, being consistent with dispersion curves in Fig. 4*B*. In the second step, we directly image Bloch waves propagating on the CROW. We select three characteristic Bloch wavevectors ($K\Lambda = 0, \frac{\pi}{2}, \pi$) as indicated by three dots in each dispersion curve in Fig. 4*B*. For the quadrupole mode in the "pole-pole" coupling configuration, as shown in Fig. 4*D*, the frequency of Bloch waves decreases as the Bloch wavevector increases. A similar situation applies to the octopole mode in the "pole-pole" coupling configuration, as shown in Fig. 4*E*. These results confirm their "backward-wave" properties.

What is more interesting is the hexapole mode that can be excited with both the "pole-pole" and "node-node" coupling configurations, where sign reversal of the coupling factor is expected. Figure 4*F* shows the measured Bloch waves of the hexapole mode in the "pole-pole" coupling configuration. It can be seen that the frequency of Bloch waves increases as the Bloch wavevector increases. This confirms the "forward-wave" properties of these Bloch waves. In Fig. 4*G*, the measured Bloch waves are plotted for the hexapole mode in the "node-node" coupling configuration. Instead, the frequency of Bloch waves decreases as the Bloch wavevector increases. Apparently, they are "backward waves."

Finally, we show that the sign-reversal coupling can occur in a single CROW, in which the Bloch waves with both positive and negative coupling factors can co-exist. We construct a right-angle bent CROW as shown in Fig. 5*A*. Similar to previous demonstrations, the end (side) excitation is achieved by placing the source and probe at positions indicated by a pair of red (blue) dots. Because of the four-fold rotational symmetry of mode profiles, the guidance of Bloch waves for the quadrupole and octopole modes will not be affected by the bending corner. In other words, their coupling factors will not change over the bending corner. On the other hand, the Bloch waves of the hexapole mode, when they propagate across the bending corner, will switch from the "pole-pole" coupling configuration to the "node-node" coupling configuration, or *vice versa*, and thus induce the sign reversal of coupling factor at the bending corner. We first measure the transmission spectra as shown in Fig. 5*B*, whose results are consistent with dispersion curves in Fig. 4*B*. We then image the Bloch waves. Both the quadrupole mode (Fig. 5*C*) and the octopole mode (Fig. 5*D*) are excited in the "pole-pole" coupling configuration. The sign-reversal coupling does not occur at the bending corner. For the hexapole mode excited with end excitation (Fig. 5*E*), the "pole-pole" ("node-node") coupling configuration applies to the horizontal (vertical) arm. The coupling factor changes its sign from negative to positive at the bending corner. The hexapole mode can also be excited with side excitation (Fig. 5*F*), where the "node-node" ("pole-pole") coupling configuration applies to the horizontal (vertical) arm. This time the coupling factor changes its sign from positive to negative at the bending corner.

In summary, by imaging the tight-binding Bloch waves on a CROW consisting of DSP resonators in the microwave regime, we have demonstrated that the coupling factor in the CROW theory can reverse its sign for a multipole resonance mode. This sign reversal of coupling factor is firstly confirmed by observing resonance wave fields in the coupled two DSP resonators. A CROW that consists of an array of DSP resonators is further constructed. By matching directly Bloch wavevectors and frequencies, it is shown that the sign-reversal coupling will induce two different waveguiding dispersion curves in the same frequency range. In view of wide applications of CROW, we expect that our study will find novel use in designs of functional photonic circuits and systems, such as in the recent topological photonics (15-18) and $\mathcal{PT}$-symmetric photonics (19-21) where systematic coupling tuning between coupled resonators is highly desirable.

**Supplementary Information** is linked to the online version of the paper at http://www.pnas.org.

**Acknowledgements** This work was sponsored by the NTU Start-Up Grants, Singapore Ministry of Education under Grant No. MOE2015-T2-1-070 and MOE2011-T3-1-005.

**Author Contributions** Z. G. and Y. Z. performed experiments. F. G. performed theoretical analysis. B. Z. supervised the project. All authors analyzed data and wrote the manuscript.

**Author Information** Reprints and permissions information is available at www.pnas.org. The authors declare no competing financial interests. Readers are welcome to comment on the online version of the paper. Correspondence and requests for materials should be addressed to B.Z. (blzhang@ntu.edu.sg).

# Figures

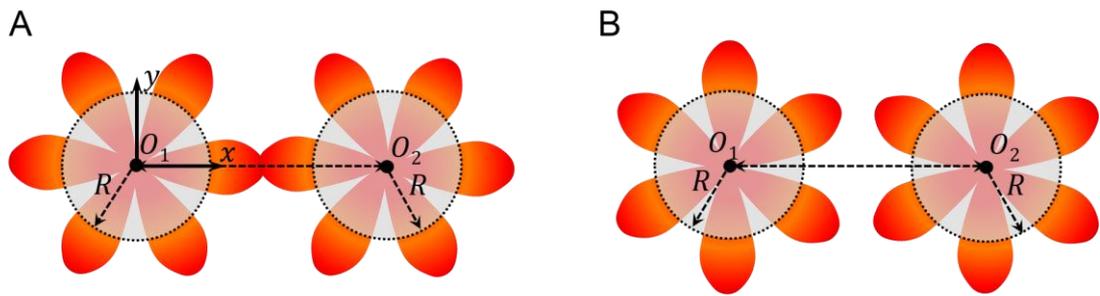

**Fig. 1.** Two coupling configurations between two coupled resonators for a multipole resonance mode. The wave pattern of multipole resonance mode is illustrated with color. (*A*) "Pole-pole" coupling configuration with overlapped intensity maxima between resonators. (*B*) "Node-node" coupling configuration with overlapped intensity minima between resonators.

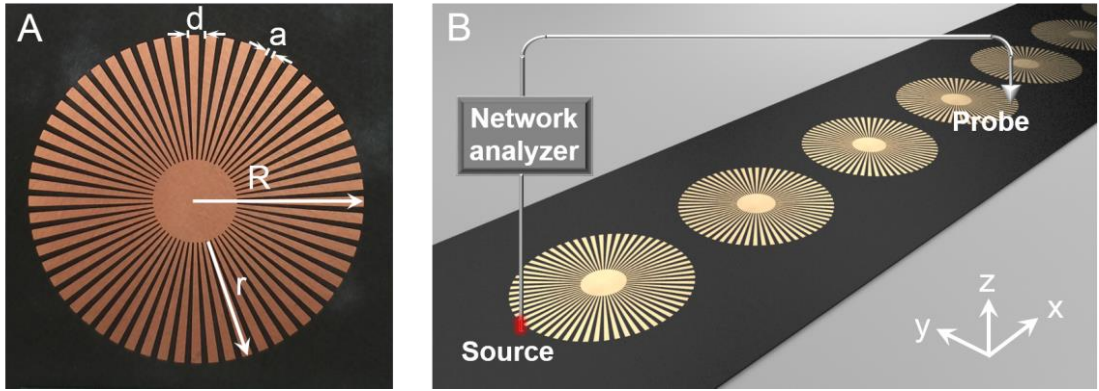

**Fig. 2.** Experimental setup for measuring coupled resonance waves. (*A*) Photo of a designer-surface-plasmon resonator with radius *R*=12 mm. The depth of grooves is *r*=9 mm. The width and periodicity of grooves are *a*=0.625 mm and *d*=1.255 mm, respectively. (*B*) Schematic of a CROW that consists of an array of designer-surface-plasmon resonators. A network analyzer records the resonance waves by scanning a near-field probe above the CROW.

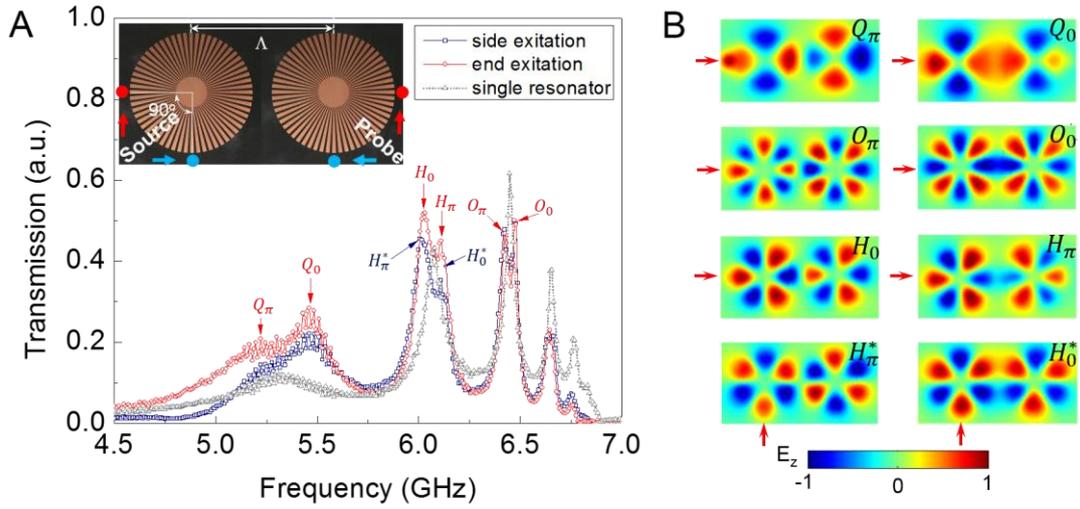

**Fig. 3.** Mode splitting of two coupled resonators. (*A*) The measured near-field transmission spectra through a coupled-resonator dimer with end and side excitations. Inset: photo of the two designer-surface-plasmon resonators with inter-resonator distance $\Lambda = 30$ mm. (*B*) The measured resonance wave patterns. The position of source is indicated by a red arrow.

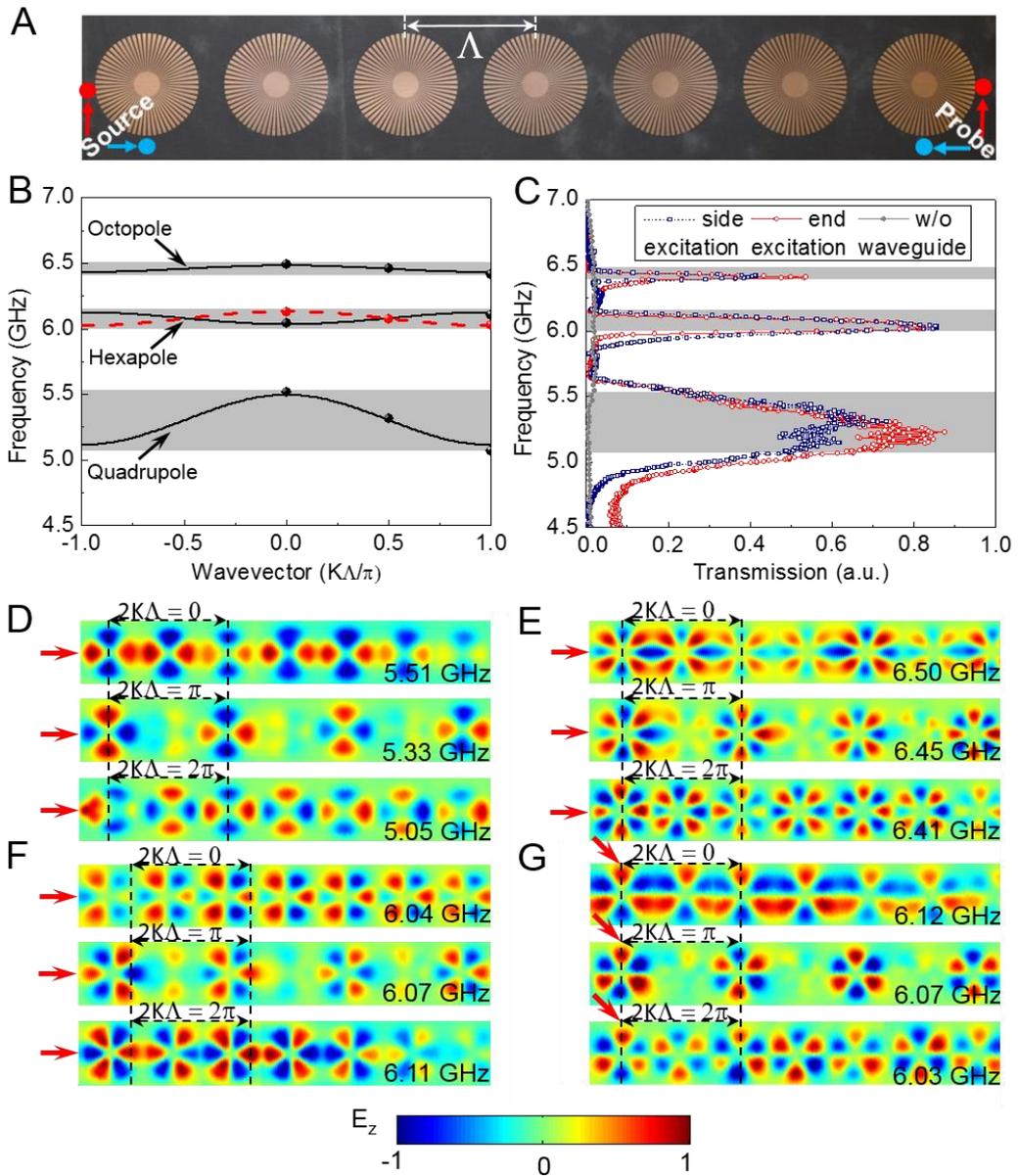

**Fig. 4.** Bloch waves on a straight CROW. (*A*) Photo of the designer-surface-plasmon CROW with periodicity $\Lambda = 30$ mm. Positions of the source and probe for the end and side excitations are indicated as a pair of red dots and blue dots, respectively. (*B*) Dispersion curves calculated with experimentally retrieved coupling factors for the quadrupole, hexapole, and octopole modes. Solid lines correspond to the "pole-pole" coupling configuration. The dashed red line for the hexapole mode corresponds to the "node-node" coupling configuration. (*C*) The measured transmission spectra through the CROW. The transmission spectrum without CROW is also measured for comparison. (*D-G*) Measured Bloch waves of the quadrupole (*D*), octopole (*E*) and hexapole (*F, G*) modes with three characteristic Bloch wavevectors as indicated by three dots in each dispersion curve in (*B*). The red arrow indicates the position of source.

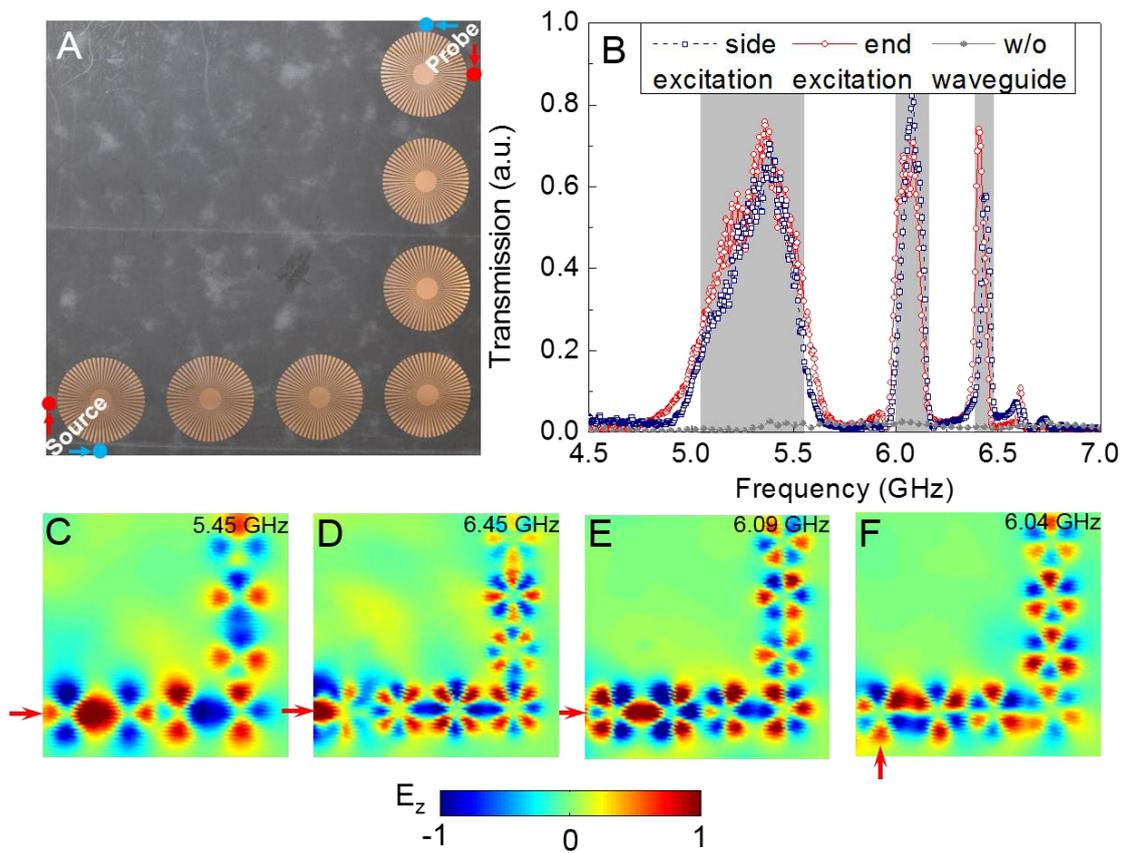

**Fig. 5.** Sign-reversal coupling in a bent CROW. (*A*) Photo of the bent designer-surface-plasmon CROW. Positions of the source and probe for the end and side excitations are indicated as a pair of red dots and blue dots, respectively. (*B*) Measured transmission spectra through the bent CROW. The transmission spectrum without CROW is also measured for comparison. (*C-F*) Measured Bloch waves of the quadrupole (*C*), octopole (*D*) and hexapole (*E, F*) modes. The red arrow indicates the position of source.

# Supplementary Information for

# "Sign-Reversal Coupling in Coupled-Resonator Optical Waveguide"

Zhen Gao[†1], Fei Gao[†1], Youming Zhang[1], and Baile Zhang*[1,2]

[1]*Division of Physics and Applied Physics, School of Physical and Mathematical Sciences, Nanyang Technological University, Singapore 637371, Singapore.*

[2]*Centre for Disruptive Photonic Technologies, Nanyang Technological University, Singapore 637371, Singapore.*

[†]*These two authors contributed equally to the work.*

[*]*Authors to whom correspondence should be addressed; E-mail: blzhang@ntu.edu.sg (B. Zhang)*

## 1) Derivation of coupling factor in "pole-pole" coupling configuration

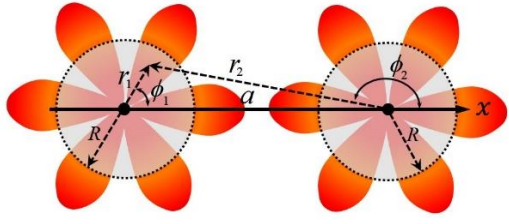

**Fig. S1.** Illustration of "pole-pole" coupling configuration.

Yariv *et al.* in Ref. 1 derived a mathematical expression for the coupling factor between adjacent dielectric resonators in a CROW as follows,

$$\kappa = \int dr^3 [\varepsilon_0(\vec{r}-a\hat{x}) - \varepsilon(\vec{r}-a\hat{x})] E_\Omega(\vec{r}) \cdot E_\Omega(\vec{r}-a\hat{x}) \qquad (1)$$

where $\varepsilon_0(\vec{r}-a\hat{x}) = \begin{cases} \varepsilon_m; & (|\vec{r}-a\hat{x}|<R) \\ \varepsilon_0; & (|\vec{r}-a\hat{x}|>R) \end{cases}$ is the dielectric constant function of a single resonator, $\varepsilon(\vec{r}-a\hat{x}) = \begin{cases} \varepsilon_m; & (|\vec{r}-na\hat{x}|<R) \\ \varepsilon_0; & \text{otherwise} \end{cases}; n \in Z$, represents the dielectric constant function of the CROW, and $E_\Omega(\vec{r})$ is the field in a single resonator with resonance frequency $\Omega$. In the following we will show the sign reversal of coupling factors for two different coupling configurations, as a supplement to discussions in the main text.

For the "pole-pole" coupling configuration as shown in Fig. S1, the resonance field profile in the left resonator, taking its center as the origin, can be written as,

$$E_1(\vec{r}) = \begin{cases} AJ_m(k_d r_1) \cdot \cos(m\phi_1) & (r_1 < R) \\ AC_m^1 \cdot H_m^{(1)}(ik_0 r_1) \cdot \cos(m\phi_1) & (r_1 > R) \end{cases}; \qquad (2)$$

where $C_m^1 = \frac{J_m(k_\rho R)}{H_m^{(1)}(ik_0 R)}$, and *A* is a constant for normalization. Similarly, the resonance field profile in the right resonator that is translated by a distance of *a* in the *x* direction can be written as:

$$E_2(\vec{r}-a\hat{x}) = \begin{cases} AJ_m(k_d r_2) \cdot \cos(m\phi_2) & (r_2 < R) \\ AC_m^1 \cdot H_m^{(1)}(ik_0 r_2) \cdot \cos(m\phi_2) & (r_2 > R) \end{cases}; \qquad (3)$$

We then substitute Eqs. (2-3) into Eq. (1). We can consider only the overlap field from the nearest resonator, because of the tight-binding nature of localized resonance modes. As a result,

$$\kappa = (\varepsilon_0 - \varepsilon_m) A^2 \int_{r_1<R} dr^3 \, C_m^1 H_m^{(1)}(ik_0 r_2) \cos(m\phi_2) J_m(k_d r_1) \cos(m\phi_1) \qquad (4)$$

Since $r_1$ and $r_2$ are with different sets of polar coordinates, we apply the addition theorem to unify them with one set of polar coordinates. This gives,

$$\kappa = (\varepsilon_0 - \varepsilon_m)A^2 \int_0^R dr_1 C_m^1 \sum_{n=-\infty}^{+\infty} H_{n-m}^{(1)}(ik_0 a) J_n(ik_0 r_1) J_m(k_d r_1) \cdot$$

$$\int_0^{2\pi} \cos(n\phi_1) \cos(m\phi_1) d\phi_1 \tag{5}$$

Only when $n = \pm m$ can the integral be nonzero. Thus,

$$\kappa = (\varepsilon_0 - \varepsilon_m)\pi A^2 \int_0^R r_1 dr_1 \{C_m^1 [H_0^{(1)}(ik_0 a) \cdot J_m(ik_0 r_1) \cdot J_m(k_d r_1) + H_{-2m}^{(1)}(ik_0 a)$$

$$\cdot J_{-m}(ik_0 r_1) \cdot J_m(k_d r_1)]\}$$

$$= (\varepsilon_0 - \varepsilon_m)\pi A^2 \int_0^R r_1 dr_1 \left\{ \frac{J_m(k_d R)}{K_m(k_0 R)} [i^{2m} K_0(k_0 a) + i^{2m} K_{2m}(k_0 a)] I_m(k_0 r_1) J_m(k_d r_1) \right\} \tag{6}$$

Since, $K_{2m}(k_0 a) \gg K_0(k_0 a)$ for tightly localized resonance modes, we can approximately obtain

$$\kappa \approx (\varepsilon_0 - \varepsilon_m)\pi A^2 \int_0^R r_1 dr_1 \cdot (i)^{2m} \cdot \frac{J_m(k_d R)}{K_m(k_0 R)} \cdot K_{2m}(k_0 a) \cdot I_m(k_0 r_1) \cdot J_m(k_d r_1) \tag{7}$$

**2) Derivation of coupling factor in "node-node" coupling configuration**

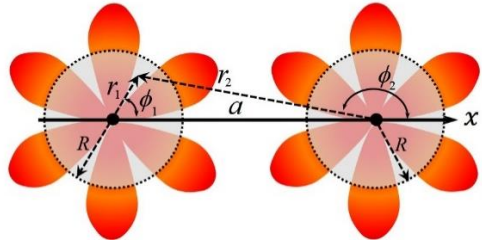

**Fig. S2.** Illustration of "node-node" coupling configuration.

For the "node-node" coupling configuration as shown in Fig. S2, the field of the left resonator can be written as:

$$E_1(\vec{r}) = \begin{cases} A J_m(k_d r_1) \cdot \sin(m\phi_1) & (r_1 < R) \\ A C_m^1 \cdot H_m^{(1)}(ik_0 r_1) \cdot \sin(m\phi_1) & (r_1 > R) \end{cases}; \tag{8}$$

The field of the right resonator can be written as:

$$E_2(\vec{r} - a\hat{x}) = \begin{cases} A J_m(k_d r_2) \cdot \sin(m\phi_2) & (r_2 < R) \\ A C_m^1 \cdot H_m^{(1)}(ik_0 r_2) \cdot \sin(m\phi_2) & (r_2 > R) \end{cases}; \tag{9}$$

Substituting Eq. (8) and Eq. (9) into Eq. (1) and performing derivations similar to last session, we can get

$$\kappa = (\varepsilon_0 - \varepsilon_m)A^2 \int_0^R dr_1 C_m^1 \sum_{n=-\infty}^{+\infty} H_{n-m}^{(1)}(ik_0 a) J_n(ik_0 r_1) J_m(k_d r_1) \cdot$$

$$\int_0^{2\pi} \sin(n\phi_1) \sin(m\phi_1) d\phi_1 \tag{10}$$

Only when $n = \pm m$ can the integral be nonzero. Thus,

$$\kappa = (\varepsilon_0 - \varepsilon_m)\pi A^2 \int_0^R r_1 dr_1 \left\{C_m^1[H_0^{(1)}(ik_0a) \cdot J_m(ik_0r_1) \cdot J_m(k_dr_1) - H_{-2m}^{(1)}(ik_0a)\right.$$
$$\left. \cdot J_{-m}(ik_0r_1) \cdot J_m(k_dr_1)]\right\}$$
$$= (\varepsilon_0 - \varepsilon_m)\pi A^2 \int_0^R r_1 dr_1 \left\{\frac{J_m(k_dR)}{K_m(k_0R)}[i^{2m}K_0(k_0a) - i^{2m}K_{2m}(k_0a)]I_m(k_0r_1)J_m(k_dr_1)\right\}$$
(11)

Since, $K_{2m}(k_0a) \gg K_0(k_0a)$ for tightly localized resonance modes, we can approximately obtain

$$\kappa \approx -(\varepsilon_0 - \varepsilon_m)\pi A^2 \int_0^R r_1 dr_1 \, (i)^{2m} \frac{J_m(k_dR)}{K_m(k_0R)} K_{2m}(k_0a)I_m(k_0r_1)J_m(k_dr_1) \quad (12)$$

Note that $\kappa$ in Eq. (12) has the same magnitude but opposite sign compared to that in Eq. (7). This shows the sign reversal of coupling factors in the two different coupling configurations.